\documentclass[preprint,authoryear,12pt]{elsarticle}
\usepackage{amssymb}
\usepackage{color}
\usepackage{amsmath}
\usepackage{subfigure}
\usepackage{graphicx}
\usepackage{graphics}
\usepackage{booktabs}
\usepackage{threeparttable}
\usepackage{appendix}
\usepackage{geometry}
\usepackage[displaymath]{lineno}
\usepackage{setspace}
\usepackage[colorlinks,linkcolor=red,anchorcolor=blue,citecolor=green]{hyperref}
\usepackage{soul}

\usepackage{mathpazo}

\journal{ }

\begin{document}
\begin{spacing}{1.5}

\begin{frontmatter}

\title{Speed Limit: Obey, or Not Obey?}

\author[rvt1]{Zhengbing~He}
\author[rvt2]{Mirco~Nanni}
\author[rvt2]{Luca~Pappalardo}
\author[rvt1]{Paolo~Santi\corref{cor1}}
\ead{psanti@mit.edu}
\author[rvt1]{Carlo~Ratti}

\address[rvt1]{Senseable City Lab, Massachusetts Institute of Technology, Cambridge MA, United States}
\address[rvt2]{Knowledge Discovery and Data Mining Lab, CNR-ISTI, Pisa, Italy}
\cortext[cor1]{Corresponding author}

\begin{abstract}
It is commonly expected that drivers maintain a driving speed that is lower than or around the posted speed limit, as failure to obey may result in safety risks and fines.
By taking randomly selected road segments as examples, this study compares the percentages of speeding vehicles in five countries worldwide, namely,
two European countries (Germany and Italy), two Asian countries (Japan and China), and one North American country (the United States).
Contrary to expectations, our results show that more than 80\% of drivers violate the posted speed limits in the studied road segments in Italy, Japan, and the United States.
In particular, a significant portion (45.3\%) of drivers in Italy exceed the posted speed limit by a substantial margin (30 km/h), while few speeding vehicles are observed in the road segment examined in China.
Meanwhile, it is found that drivers on low-speed-limit roads are more likely to exceed the posted speed limit, particularly when there are fewer on-road vehicles.
The comparison of different countries' speeding fines indicates that for the purpose of preventing speeding, increasing fines (as Italy has done) is less effective than enhancing supervision (as China has done).
The findings remind law enforcement agencies and traffic authorities of the importance of the supervision of driver's behavior and the necessity of revisiting the rationale for the current speed limit settings.

\end{abstract}


\end{frontmatter}




\newpage

\section{Introduction}

Setting an adequate road traffic speed is fundamental for human mobility.
However, the pursuit of human beings to shorten on-road travel time brings a variety of high speed-related issues, such as unsafe driving, abundant emissions, and high noise levels \citep{Nilsson1981, AARTS2006, Barth2008}.
According to the World Health Organization, approximately 1.25 million people worldwide lose their lives in road traffic accidents every year, and a third of these accidents are related to high driving speeds \citep{WHO2017}.
Due to realizing the severity of the issues, the United Nations has incorporated reducing global deaths and injuries from road traffic accidents (Target 3.6) and creating safe transportation systems (Target 11.2) in their Sustainable Development Goals (SDG) to urge countries to better regulate vehicle speed.

Establishing speed limits is the most important and common approach to controlling vehicle speed.
Unfortunately, the number of speeding-related accidents remains high despite speed limit signs being posted everywhere.
For example, 28\% of fatal crashes and 13\% of injury crashes in 2020 were speeding-related in the United States \citep{NHTSA2020}.
In 2020, inappropriate speed was involved in approximately 13.9\% of injury crashes and 33.7\% of fatal crashes in Germany, while speed limit violations were recorded in 10\% of all road fatalities in Italy \citep{ITF2021}.

All of these accident-related statistics reflect the consequences of speeding, but they do not serve as direct knowledge of drivers' speeding behavior.
To understand speeding behavior itself, questionnaire surveys are the prevailing method and are usually employed to understand traffic participants' thoughts and habits.
Road safety performance indicator-related personal data were collected through roadside surveys in 29 European countries between 2004 and 2007 \citep{Vis2007}.
An online questionnaire study was conducted across 32 countries, and risky driving behaviors were investigated, including alcohol/drug driving, speeding, and mobile phone use while driving \citep{PIRES2020}.
A questionnaire survey was also used to investigate drivers' long-term changes in their compliance with speed limits after changing speed limits and increasing speed enforcement and public education \citep{Son2009, VANKOV202110}.

Traffic speed regulations are often associated with sanctions and the same level of law enforcement.
Currently, almost all countries have specific articles of law to avoid drivers' violations of speed limits, and the related sanctions mainly include monetary fines, penalty point accumulation, and even revoking driver's licenses.
For example, Italy is known for high monetary fines for speeding, while the fines in China are relatively lenient \citep{Zhao2019}.
The levels of law enforcement are also different across countries.
In many countries, such as the United States, the primary means of traffic law enforcement is that police officers are randomly or selectively stationed on roadsides and issue speeding drivers traffic tickets \citep{LIU2019}.
This results in a relatively low-level means of law enforcement because of the low spatio-temporal coverage that results from the limited number of patrolling police officers.
In contrast, in some countries, such as China and India, traffic is heavily surveilled \citep{Statista2022}, making
it more probable for violations to be caught.
Regarding the effectiveness of law sanctions and enforcement, some before-and-after comparative studies indicated that increasing the level of law enforcement could reduce the number of speed limit violations \citep{Simpson2020, MARTINEZRUIZ2019267, WALTER20111219}, while the results of a questionnaire study showed that the effect of increasing sanctions is relatively marginal \citep{RYENG2012446}.

Most of the existing understanding of speeding behavior comes from questionnaire-based stated preference surveys.
Stated preference surveys are suitable for understanding traffic participants' personal attitudes, while self-reported results may be biased by people's real-world behavior \citep{PETERSON2021}.
Moreover, the results are an indirect reflection of real traffic and still cannot give an estimation of the real number of drivers who violate the speed limit.
Therefore, direct observation and an understanding of the (speeding) vehicles operating on roads are still lacking.

In addition to the direct observation of real-world traffic, cross-country comparisons would provide good opportunities to learn lessons in policy-making and policy-implementing from other countries, with no need for making imaginary scenarios or practically modifying laws.
Such comparisons comprise a benefit that cannot be obtained from most of the existing studies, which only spotlight a single country, such as the United States \citep{PETERSON2021}, Colombia \citep{MARTINEZRUIZ2019267}, Malaysia \citep{Mohamad2019}, Sweden \citep{VADEBY201834}, or Japan \citep{Dias2018}.

In summary, it is known from statistics on traffic accidents and the results of a variety of questionnaire surveys that speeding is a dangerous but difficult-to-control behavior.
However, empirical knowledge based on the direct observation of speeding in real-world traffic is rare.
This knowledge is crucial because it enables us to clearly assess the rationality of the current speed limits and further clarify the bottleneck in the speed management chain represented by ``speed limit setting $\rightarrow$ law enforcement $\rightarrow$ law penalty".
To fill this gap, this paper analyzes traffic data collected from 5 randomly selected road segments in two European countries (Germany and Italy), two Asian countries (China and Japan), and one North American country (the United States) and compares the speeding behavior in the examined segments and the associated law enforcement and penalties.
To the best of our knowledge, the following findings are obtained from real-world traffic observations for the first time.
\begin{itemize}
\item
More than 80\% of drivers violate the posted speed limits in the randomly selected road segments in Italy, Japan, and the United States.
In particular, a significant portion (45\%) of drivers in Italy exceed the posted speed limit by a substantial margin (30 km/h), while few speeding vehicles are observed in the road segment examined in China (9.9\%).

\item
When the traffic is lighter, low speed limits are more likely to result in speed limit violations, as seen by comparing the traffic in the road segment in Japan with those in Germany and the United States.

\item
Solely increasing the severity of law penalties may not reduce speed limit violations (as shown in the case of Italy). In contrast, increasing the level of law enforcement might be more effective (as shown in the case of China).

\end{itemize}

The organization of this paper is as follows:
Section \ref{sec_result} presents the main findings and results, including the percentages of speeding drivers from the compared road segments in the five examined countries, the comparison of the laws and law enforcement, etc.
Section \ref{sec_discussion} discusses the results.
Section \ref{sec_method} introduces the method of splitting free-flow and congested conditions of traffic.
Sections \ref{sec_data} and \ref{sec_Law} present the data and relevant laws of speed limit violation in the five examined countries, respectively.

\section{Results}\label{sec_result}

This study compares the percentages of speeding vehicles in five randomly selected road segments in five countries around the world, namely, Germany, Japan, China, Italy, and the United States (see Section \ref{sec_data} for the data).
The percentages at four levels of speeding are examined, which are exceeding the speed limit, exceeding the speed limit over 10 km/h, exceeding the speed limit over 20 km/h, and exceeding the speed limit over 30 km/h.
The levels of speeding use the absolute magnitude of speed to align with the settings of the law penalty for speeding in most countries (see Section \ref{sec_Law} for law enforcement).

\begin{figure}[h]
    \centering
    \includegraphics[width=0.85\textwidth]{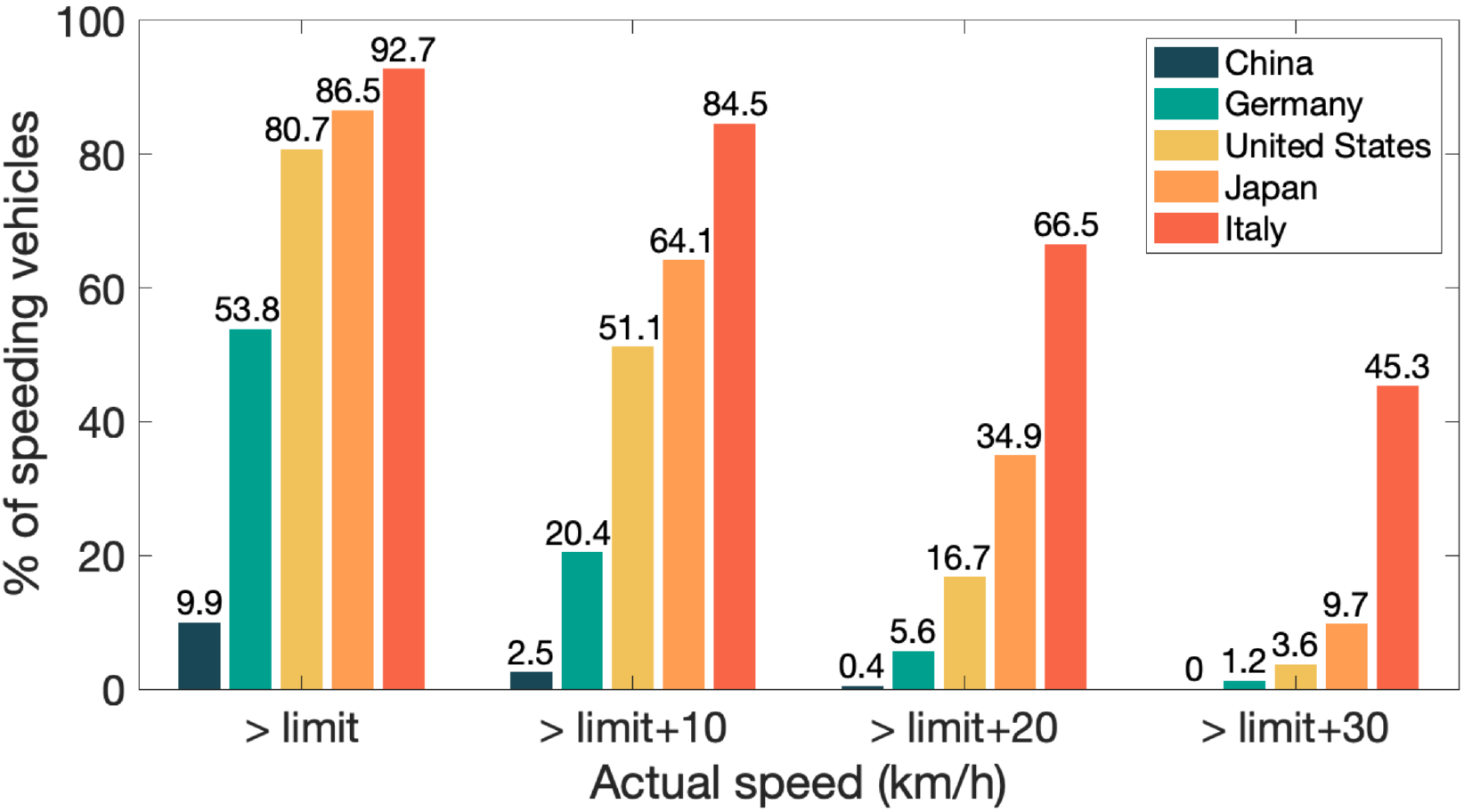}
    \caption{Comparisons of the percentage of speeding vehicles in China, Germany, Japan, United States, and Italy.}
    \label{fig_Result1}
\end{figure}

The result in Figure \ref{fig_Result1} shows that in free-flow traffic (extracted and validated in Section \ref{sec_method}), where it is widely believed that individual drivers have the freedom to choose their driving speed, more than 80\% of drivers exceed the posted speed limits in the road segments in Italy, Japan, and the United States.
The fewest speeding drivers are observed in the road segment in China, i.e., only approximately 9.9\%.
The percentages of speeding drivers in the road segment in Italy are the highest found in all levels of speeding; in particular, for the maximum level of 30 km/h, the percentage in Italy reaches 45.3\%.
The results imply that speeding seems to be quite a common behavior in many countries, such as Germany, Japan, Italy, and the United States.
This finding is surprisingly contrary to common expectations; i.e., obeying speed limits is a fundamental rule prescribed by law and thus, the provisions of the law should be respected and abided by every citizen.

\begin{figure}[h]
    \centering
    \subfigure[Germany (speed limit=120 km/h)]{
    \includegraphics[width=3in]{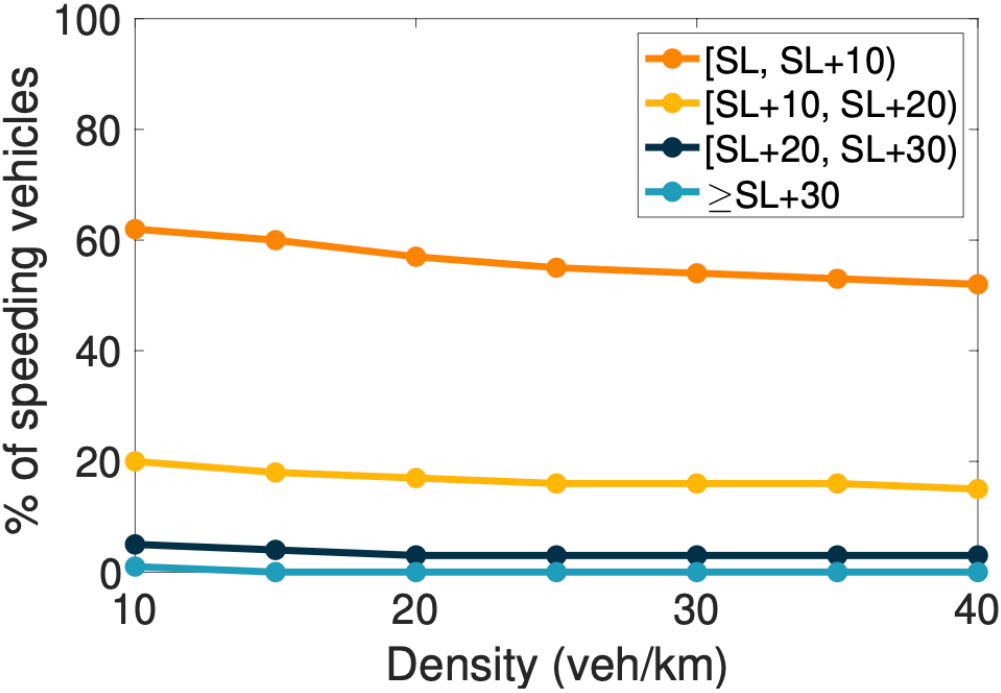}}
    \subfigure[United States (speed limit=98 km/h)]{
    \includegraphics[width=3in]{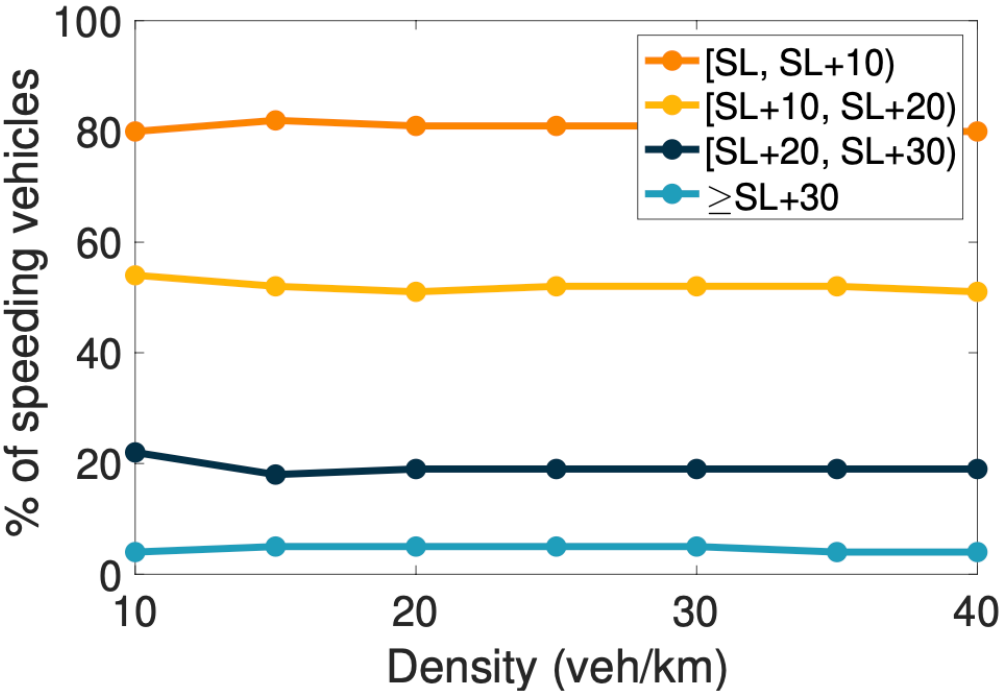}}    
    \subfigure[Japan (speed limit=60 km/h)]{
    \includegraphics[width=3in]{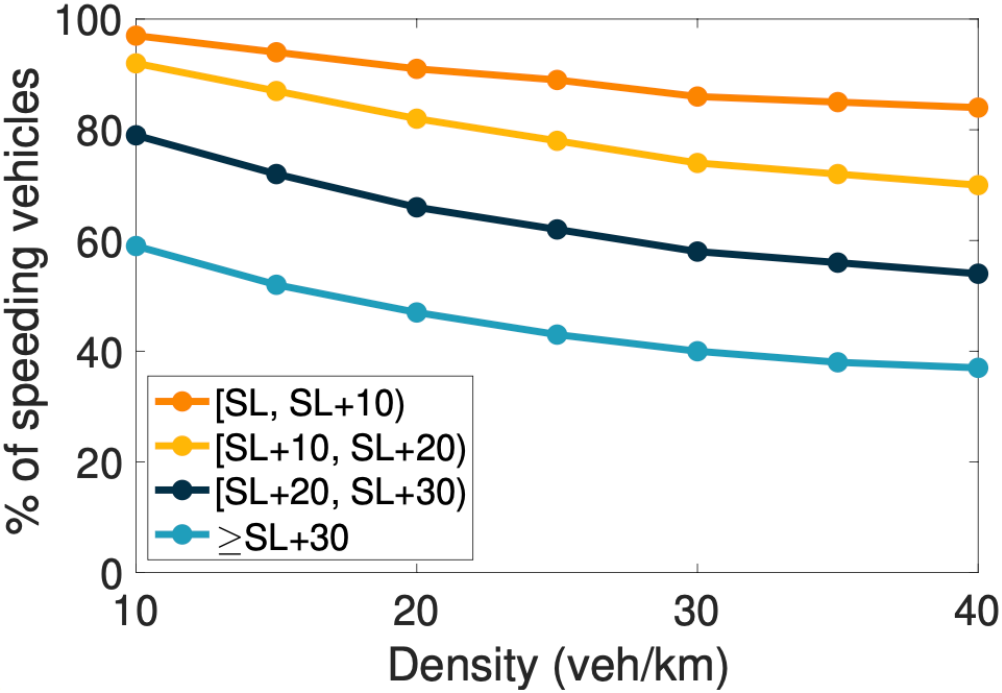}}
    \caption{Percentages of speeding vehicles associated with traffic density.}
    \label{fig_Result2}
\end{figure}

Two different reactions of drivers to the changes in the number of free-flowing vehicles are observed in the road segments in Japan, Germany, and the United States\footnote{As described in Section \ref{sec_data}, only the datasets of Japan, Germany, and the United States contain headway data, allowing us to analyze the density-related relations.}:
(i) The percentage of speeding drivers in the road segment in Japan (where the speed limit is 60 km/h) increases with a decrease in the number of on-road vehicles (Figure \ref{fig_Result2}(a)).
This inversely proportional relationship is consistent with the result of \cite{MARSHALL2023107038}, where light traffic was analyzed under stay-at-home orders during the COVID-19 pandemic.
(ii) The number of on-road vehicles has no clear impact on the percentage of speeding drivers in the road segments in Germany and the United States (where the speed limit is 120 km/h and 98 km/h, respectively); i.e., the inversely proportional relationship seen in (i) is not clearly observed (Figures \ref{fig_Result2}(b) and \ref{fig_Result2}(c)).
The difference in speed limits might account for the difference in the relationship between speeding driver percentages and vehicle numbers.
More specifically, willingness and capability might be two of the factors that make people drive faster.
The willingness to drive faster generally increases when there are fewer on-road vehicles.
However, people have the capability to drive faster only when the speed limit is low.
This implies that drivers on low-speed-limit roads are more likely to exceed the speed limit, particularly when there are fewer on-road vehicles.

\begin{figure}[h]
    \centering
    \includegraphics[width=0.85\textwidth]{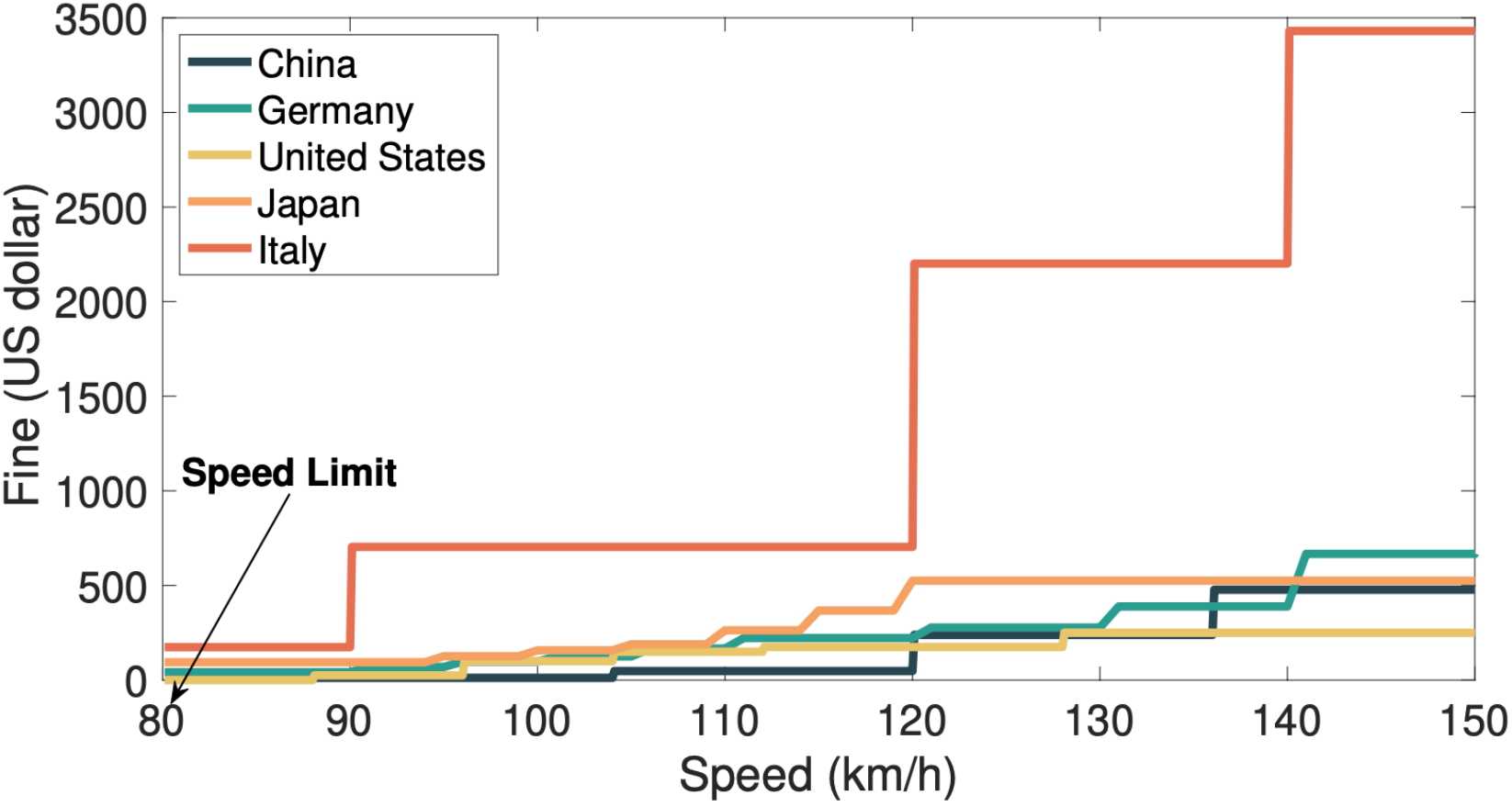}
\caption{Comparison of fines for speeding in China, Germany, Japan, the United States, and Italy, under the assumption that the speed limit is 80 km/h.}
    \label{fig_Fine}
\end{figure}

A follow-up question arises, namely, what factors result in the notable disparity in speeding behaviors between the road segments in Italy and China (Figure \ref{fig_Result1})?
To answer this question, a survey on the speeding-related provisions of laws in those five countries is conducted in Section \ref{sec_Law}.
Generally, there are two simultaneous ways of punishing speed limit violations all around the world, namely, monetary fines and penalty-point accumulation.
By taking monetary fines as an example, the severity of the penalty for speed limit violations is compared in the five countries (Figure \ref{fig_Fine}).
To make the plots more comparable rather than using simple exchange rates, the local currency was converted to US dollars using purchasing power parity\footnote{Converted on June 1st, 2023 using the converter in \url{https://www.chrislross.com/PPPConverter/}} (PPP), which is the currency conversion rate considering purchasing power, inhabitant income level, etc.
Surprisingly, while the penalty for speeding in Italy (i.e., the green line in Figure \ref{fig_Fine}) is much heavier than those in other countries, more speeding vehicles were found in the Italian road segment.
In contrast, while the fines in China are quite light (i.e., the red line in Figure \ref{fig_Fine}), fewer speeding vehicles were found.
This twist, on the one hand, reflects the intention of the Italian government to regulate speed limit violations.
On the other hand, it shows the failure of simply increasing fines.

What results in the Chinese success in speeding control based on this studied segment, given the country's lighter monetary fines?
The answer might be found in the high level of law enforcement brought about by the large number of surveillance cameras that almost cover every corner of a city \citep{Ani2023}.
In the most surveilled country worldwide, every single violation could be caught by the ``invisible" eyes.
Being caught speeding by 10\%-20\% 4 times (or by 20\%-30\% twice) within a year will result in the loss of all points and the suspension of one's license (Table \ref{fig_Law}).
Leaving the concern of personal privacy and freedom aside, such a dense surveillance system makes the penalty-point accumulation system play a more important role in speed control than monetary fines and thus effectively prevents large-scale speeding.
In contrast, the spatiotemporally random patrol of a limited number of police officers in other countries does not provide enough deterrents to prevent speeding.

Laws are comprised of a set of rules that are created and enforceable by social or governmental institutions to regulate behavior, guaranteeing the functioning of society \citep{Robertson2013}.
Laws take effect only when sufficient supervision is available.
When law enforcement agencies cannot make sufficient supervision, a heavy penalty is usually imposed with the expectation of such a penalty being an effective deterrent for offenders.
This approach may work particularly when people can feel that there is a high potential risk of being caught engaging in lawbreaking behavior themselves.
However, this approach might not be effective when people believe that there is only a small probability of them being caught engaging in lawbreaking behavior or that the penalty if they are caught will not hurt them, such as the speed limit violations that are discussed in this paper.
Realizing the current difficulty of monitoring on-road vehicles, the European Union launched the Intelligent Speed Adaptation Program in 2022.
The anti-speeding technology can sense vehicle locations using GPS devices and alert speeding drivers via an in-vehicle warning system. 
This program provides a good opportunity for Europe to better supervise driving behavior at a low cost.

Although the law should not be broken under any circumstances, the observation of a large proportion of speed limit violations, particularly given the large magnitude of such violations (such as those seen in the road segment in Italy), reminds traffic authorities and engineers to re-examine the current posted speed limits.
As mentioned in \cite{Pinna2020}, solely considering geometric characteristics and road infrastructure, instead of drivers' behavior, makes people often drive at a speed higher than the design speed in Italy.
Therefore, historical naturalistic driving data and crash data are the important basis of such examination.
If few crashes have occurred over the past years, it might be rational to increase the speed limit to better align with drivers' behavior and desires because drivers have shown that the road conditions allow them to engage in higher driving speeds.
If it is definitely necessary to implement the current speed limit owing to other considerations, then intense law enforcement is needed to ensure the effectiveness of the speed limit.

Finally, we respond to the question presented in the title of the paper in the following one sentence:
{\it While laws in countries include provisions that stipulate drivers obey the posted speed limits, many drivers who are law-abiding citizens do not obey these speed limits; thus, increasing the level of law enforcement might be the cure for speeding instead of simply increasing the related penalties.}

\section{Discussion}\label{sec_discussion}

The paper attempts to learn knowledge and lessons about speed limit violations by comparing the direct observation of road traffic in different countries.
To date, it is impossible to make a large-scale collection of data across countries; thus, only five randomly selected road segments in the studied countries are examined.
We expect the pilot-style study described herein to provide useful insight into speeding behavior and the associated law and law enforcement in different countries instead of the wishful thinking reflected by the overall situations within those countries.
Therefore, the possibility that high or low percentages of speed limit violations only occur in the sampled road segments in those countries cannot be ruled out.

Regarding the case of Italy, we also investigate GPS data from the road segments around the studied road segment; the results are consistent with those reported in the paper, thereby confirming the severe and large proportion of speeding in that area.
Most likely due to certain considerations, the speed limit for the secondary roads in that area was set to 50 km/h, while the common limit for secondary roads in Italy is 90 km/h.
This local inconsistency might be one of the factors that result in the current speeding situation, which reminds us of the importance of speed limit consistency.

As Theodore Roosevelt proclaimed, ``{\it No man is above the law and no man is below it}"; drivers should follow the posted speed limits regardless of road grades.
In this regard, it is not necessary to classify roads according to their grades.
However, the speeding behavior in practice might be different; i.e., the percentages and magnitudes of speeding might vary across road grades.
Therefore, considering road grades could be a future research direction.
Likewise, expanding the size of the samples to include more countries, more roads, and more grades of roads is a straightforward future direction that could provide more useful information to deepen and widen our understanding.

According to the results of this paper, revisiting the current setting of speed limits is necessary and crucial for the better control of speed limit violations.
In the era of big data, traditional manpower-based methods could be replaced or improved by various data-driven methods, such as taking advantage of floating car data to understand driver behaviors and Google Street View data to examine road environments.
The new types of data facilitate the semi-automatic or even automatic evaluation of the setting of speed limits, thus making it both possible and feasible to conduct an efficient and effective citywide or even nationwide speed limit examination.

Moreover, it would be interesting to know if the daily disregard-for-law behavior of speeding undermines the authority and seriousness of the entire legal system, which would represent a contagion effect across different aspects of abiding by the law.
Thus, it is possible that such an effect exists but has not been empirically confirmed yet.

Last but not least, as law enforcement plays a critical role in speeding control, it is urgent to find the optimized level of law enforcement for speeding control.
There might be a diminishing-marginal-utility relationship between the number of surveillance units (policy or camera) and the effectiveness of speeding control; i.e., the effectiveness of increasing the number of surveillance units might be quite obvious at the beginning, while it will become gradually marginal with the growth of the number of units.
Therefore, it is possible to find an optimized number of surveillance units that could balance both investment and effectiveness.

\section{Method}\label{sec_method}

Drivers have chances to exceed speed limits only in free-flow conditions \citep{TRB2011, RICHARD2020}.
Therefore, the extraction of vehicles operating under free-flow conditions is the main goal of the data preparation conducted prior to formal analyses.

\subsection{Distinguishing free-flow conditions given 100\% vehicle trajectories}\label{sec_method_trajectory}

Generally, there is no clear boundary (i.e., critical density) that could explicitly split free-flow and congested traffic conditions.
According to \cite{TRB2010} (e.g., Exhibits 11-2 and 14-5), the critical density is approximately 28 veh/km (i.e., 45 pc/mi/ln in \cite{TRB2010}) when the free-flow speed is 72$\sim$98 km/h (i.e., 45$\sim$ 60 mi/h in \cite{TRB2010}).
The results in Figure \ref{fig_Result1} (i.e., the percentages of speeding drivers) are not quite sensitive to the critical density, particularly between 20$\sim$40 veh/km, as shown in Figure \ref{fig_Result2}.
Therefore, the critical density that is used to split free-flow and congested traffic conditions is selected to be 30 veh/km in the current paper.

Given 100\% high-fidelity vehicle trajectories, the speed and spacing headway can be directly calculated for each vehicle.
Only the moment when the maximum speed occurs in the studied road segments is considered, as the maximum speed is more related to speed limit violations.
Therefore, the free-flow conditions for an individual vehicle can be determined as follows:
A vehicle is considered to be free-flowing if $s'>s^*$,
where $s'$ is the spacing headway when the maximum speed occurs, and $s^*$ is the critical spacing headway converted from the critical density $k^*$ as $s^*= 1/k^*$, where $k^*=30$ veh/km, as mentioned above.

\subsection{Inferring free-flow conditions given only speed data}\label{sec_method_normal}

Strictly speaking, it is impossible to confirm free-flow conditions solely based on partially existing traffic speed data collected by, e.g., probe vehicles \citep{Ambros2020}.
The distinguishing feature of free-flow traffic is that there is no mutual interaction among vehicles.
Therefore, we can infer from a statistical perspective that vehicles are in free-flow conditions when their speeds adhere to a normal distribution, considering the fact that drivers can independently choose their speeds in free-flow conditions. The one-sample Kolmogorov-Smirnov (KS) test \citep{Massey1951} is employed to infer if a group of speed data follows a normal distribution.

\section{Data}\label{sec_data}

\subsection{Description of the data}

\subsubsection{Germany}

The highD dataset\footnote{The highD dataset found at \url{https://www.highd-dataset.com}} is a dataset of naturalistic vehicle trajectories recorded on German highways using drones \citep{highD} was employed to investigate speeding behavior in Germany.
The trajectory data from Records 31$\sim$50 in the dataset, which were collected from Segment 1 (51.1163$^\circ$ N, 6.5886$^\circ$ W) on Freeway A46, were used in the current paper.
The average duration of the records is 17 min, and the length of the freeway segment is 420 m.
The segment contains two directions, and there are three lanes in each direction with the labeled numbers of 2$\sim$4 and 6$\sim$8.
Most vehicles in Lane 8 were trucks, making the traffic different from that in the other lanes.
Thus, only the data from the other Lanes 6 and 7 in one direction are used herein.
The speed limit was 120 km/h.

\subsubsection{Japan}

The ZenTraffic dataset\footnote{The ZenTraffic dataset found at \url{https://www.zen-traffic-data.net}} is a dataset of naturalistic vehicle trajectories recorded on Japanese expressways using roadside cameras \citep{ZenTraffic} was employed to investigate speeding behavior in Japan.
The trajectory data from Records L001\_F001$\sim$F005, which were collected from Hanshin Expressway Route 11 (34.7151$^\circ$ N, 135.4656$^\circ$ W), were used in the paper.
Each record lasts for 60 min, and the length of the expressway segment is 2000 m.
The dataset only contains the vehicle trajectories in one direction with two lanes (the Osaka-bound Ikeda Route), and all the data are involved.
The speed limit was 60 km/h.

\subsubsection{United States}

The traffic data in the United States come from the CitySim dataset\footnote{The CitySim dataset found at \url{https://github.com/ozheng1993/UCF-SST-CitySim-Dataset}} is a dataset of naturalistic vehicle trajectories collected using drones \citep{Zheng2022}.
The data collection location was a freeway segment in Florida.
The record consists of three pieces with a total of approximately 38 min, and the length of the covered segment is approximately 140 m.
It is observed from the videos that the number of trucks is low; thus, all data from one 5-untolled-lane direction are used.
The speed limit was 98 km/h (i.e., 60 mi/h).

\subsubsection{Italy}
The traffic data in Italy were collected by onboard units instrumented in vehicles for the purpose of insurance-related driving monitoring.
The median value of the sampling interval is approximately 90 sec, while the penetration rate is unclear.
One-day data were collected on an 800-m segment of an Italian secondary road (SS16) in the city of Ravenna (44.3821$^\circ$ N, 12.2014$^\circ$ W).
The vehicles were randomly sampled, making the sampled data representative.
The speed limit was 50 km/h.

\subsubsection{China}
The data from China were collected from the regularly reported speed and position data from taxes \citep{He2017}.
The location of collection was a 1000-m urban freeway segment of the east 3rd Ring Road of Beijing (39.9352$^\circ$ N, 116.3101$^\circ$ W), and the sampling interval was approximately 1 min.
Only the data gathered from 11:00 PM to 6:00 AM during the whole month of January 2015 were employed considering the fact that traffic at midnight is commonly uncongested and drivers tend to drive faster \citep{BASSANI2016, Ambros2020}.
The speed data from taxes are able to reflect the traffic conditions surrounding the taxes according to the rule of ``follow the flow"; i.e., vehicles usually move within the traffic instead of obviously slower or faster than the others.
The speed limit was 80 km/h.

\subsection{Statistics of the data}

\begin{figure}[!htbp]
    \centering
    \subfigure[Japan]{
    \includegraphics[width=3in]{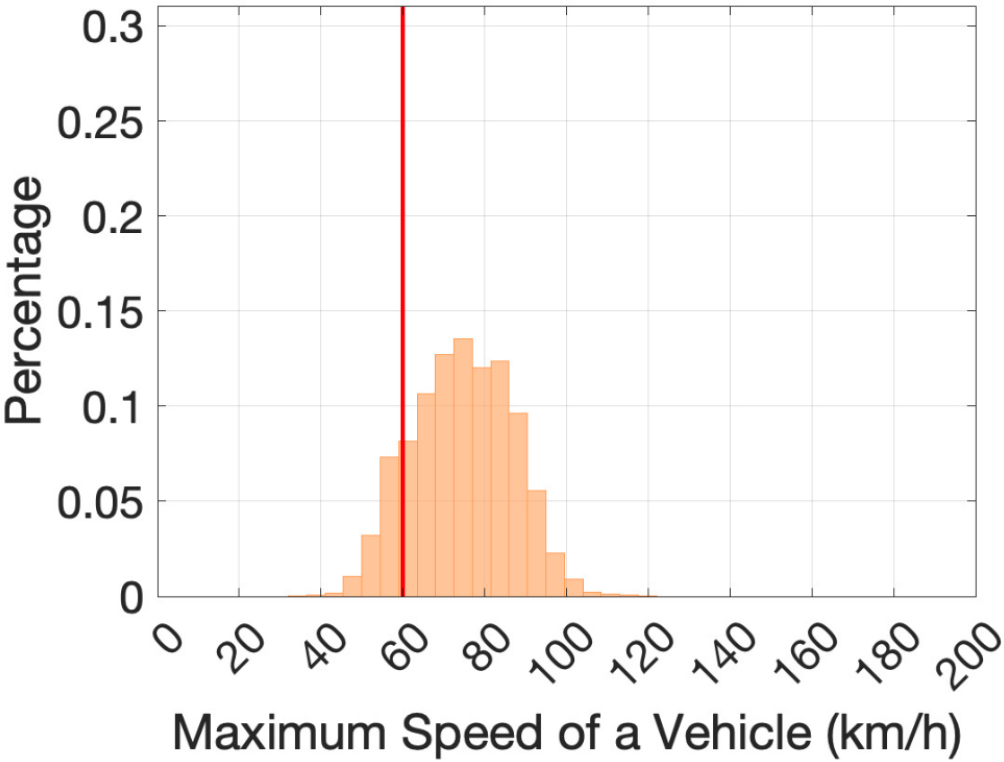}}
    \subfigure[China]{
    \includegraphics[width=3in]{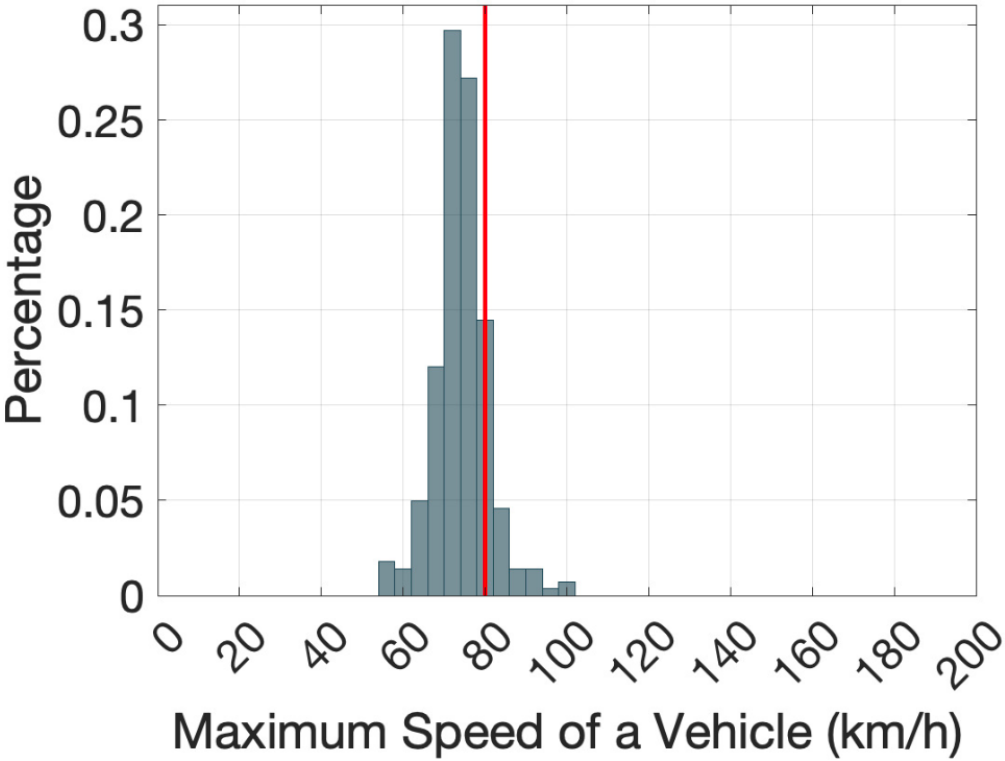}}          
    \subfigure[Germany]{
    \includegraphics[width=3in]{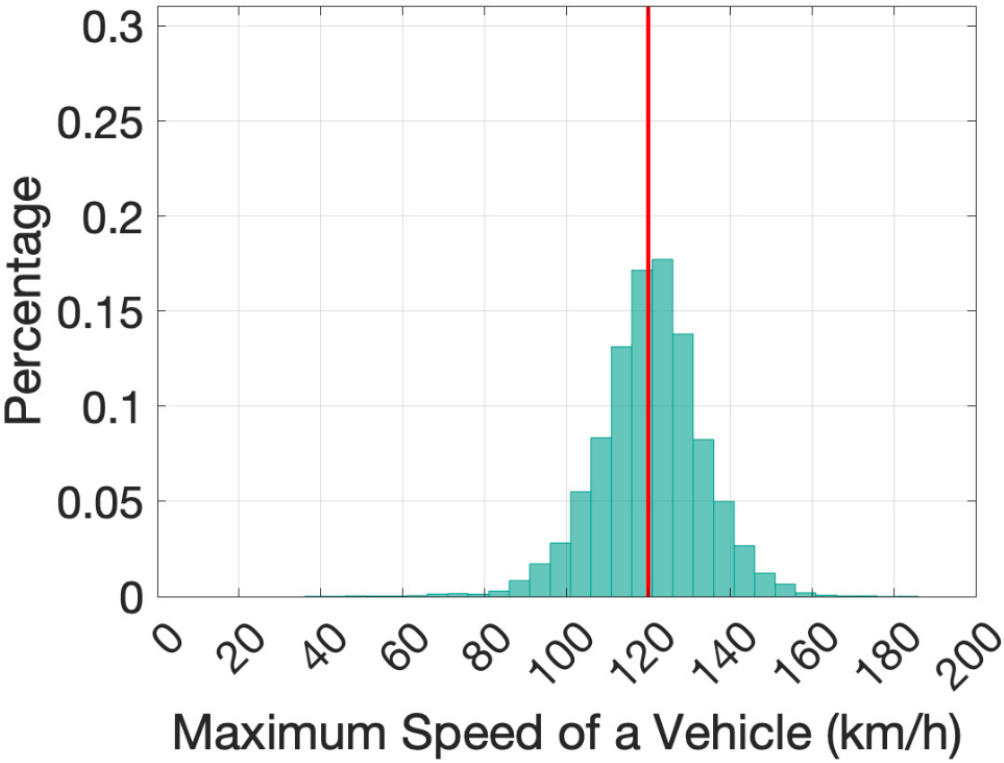}}
    \subfigure[Italy]{
    \includegraphics[width=3in]{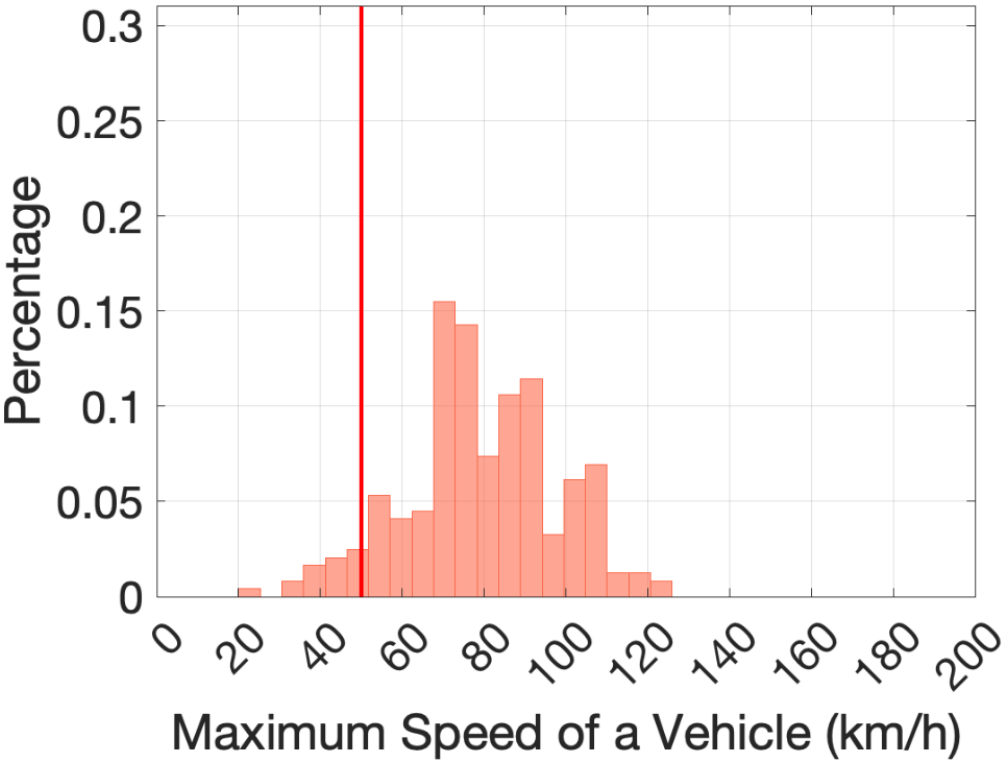}}
    \subfigure[United States]{
    \includegraphics[width=3in]{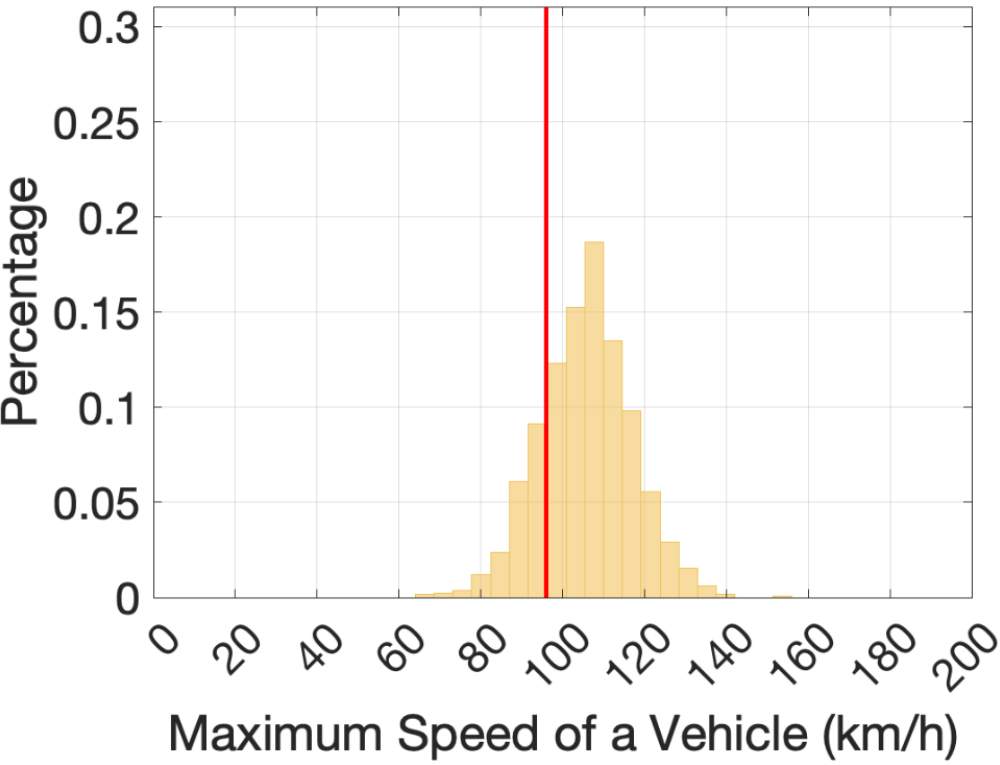}}          
\caption{Distributions of the free-flowing vehicles in the road segments in Japan, China, Germany, Italy, and the United States. The red lines indicate the speed limits.}
    \label{fig_Dist}
\end{figure}

The data used herein are summarized in Table \ref{tab_Data}.
According to Section \ref{sec_method_trajectory}, the maximum speed of each vehicle follows a normal distribution (Figure \ref{fig_Dist}), as supported by the KS test with a significance level of 0.05. This confirms that the speed is the result of drivers' independent selections; i.e., the vehicles were in free-flow conditions.

\begin{table}[!htbp]\centering\footnotesize
\caption{Summary of the data collected from Japan, China, Germany, Italy, and the United States.}\label{tab_Data}
\setlength\tabcolsep{2pt}
\begin{tabular}{llllllll}
\toprule
  Country & Data & Road & Lane   & Speed Limit    & Vehicle  & Year of   & KS Test for\\
          & Type & Type & Number & (km/h)         & Number   & Collection  & Normal Dist.$^{[1]}$  \vspace{1mm}\\
\midrule
  Germany & Traj (100\%)    & Freeway           & 3   & 120         & 12897  & 2017-2018 & Passed  \vspace{1mm}\\
  Japan   & Traj (100\%)    & Expressway        & 2   & 60          & 9053   & 2018      & Passed  \vspace{1mm}\\
  U.S.    & Traj (100\%)    & Freeway           & 5   & 98$^{[2]}$  & 1810   & 2022      & Passed  \vspace{1mm}\\
  Italy   & GPS (samples)   & Secondary road    & 2   & 50          & 245    & 2019      & Passed \vspace{1mm}\\
  China   & GPS (samples)   & Urban freeway     & 3   & 80          & 283    & 2015      & Passed \\
\bottomrule
\end{tabular}
\begin{tablenotes}\footnotesize
\item[1] \hspace{-4mm} [1] The significant level is 0.05.
\item[2] \hspace{-4mm} [2] 60 mile/h.
\end{tablenotes}   
\end{table}

\section{Laws and penalties on speed limit violation}\label{sec_Law}

\begin{figure}[!htbp]
    \centering
    \includegraphics[width=1\textwidth]{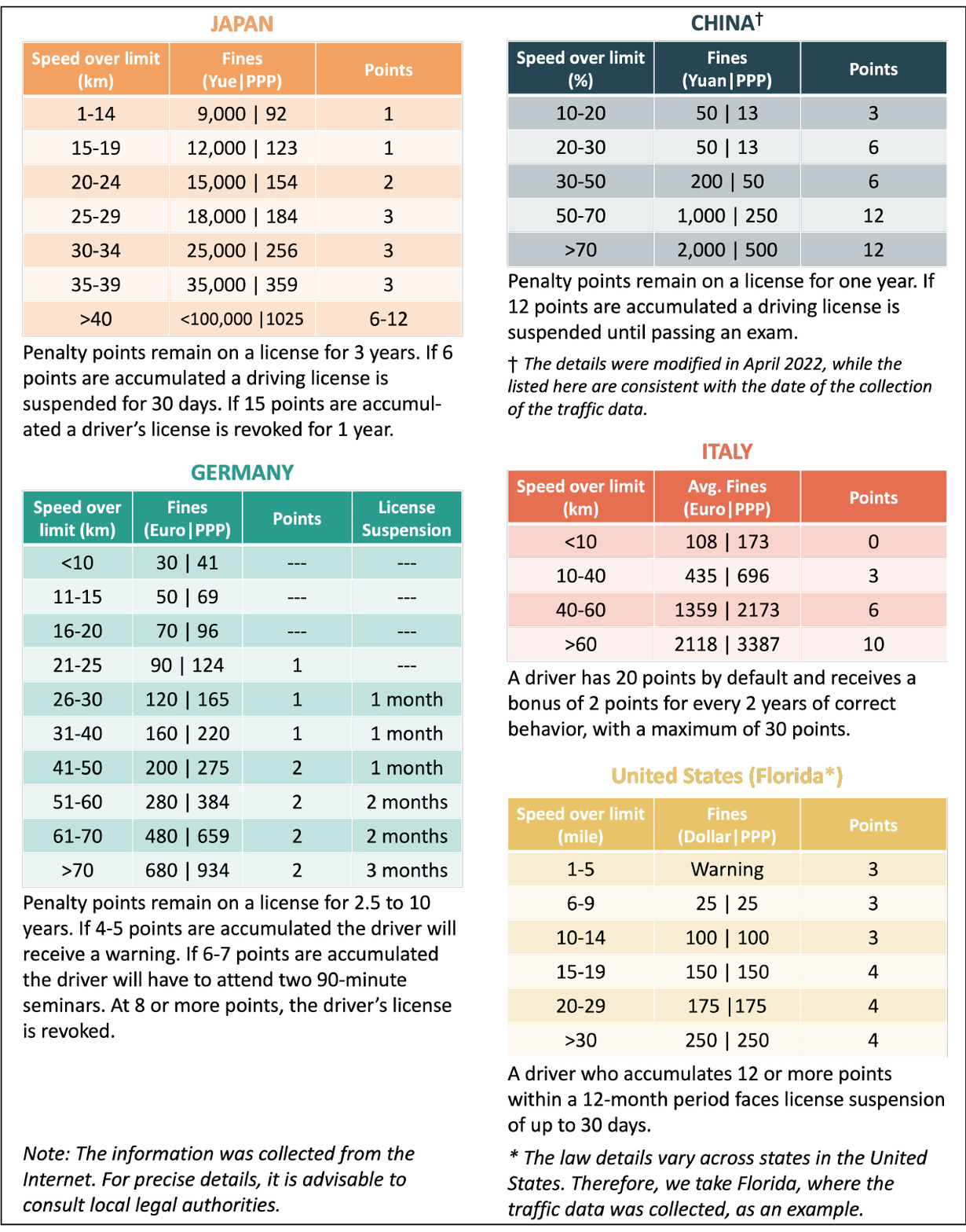}
\caption{The penalties for speed limit violations in Japan, China, Germany, Italy, and the United States.{\bf PS comment: to improve comparison, shouldn't we report the dollar equivalent for the fines?}}
    \label{fig_Law}
\end{figure}

Monetary fines and penalty-point accumulation are the major ways of punishing speeding violations all around the world.
\begin{itemize}

\item \textbf{Monetary fines}. The amount of the fine typically depends on the severity of the speed limit violation and is determined by the relevant laws and regulations of each country.

\item \textbf{Penalty-point accumulation}. Each violation carries a specific number of penalty points, which vary based on the degree of speeding. When a driver accumulates a certain number of penalty points within a specified period, it can result in additional consequences such as license suspension, mandatory driver improvement courses, or other penalties.

\end{itemize}
The laws and penalties from the five countries are summarized in Figure \ref{fig_Law} for the purpose of comparison.

\section*{Acknowledgments} 
The authors are grateful to Ashutosh Kumar for processing video data, to Dr. Jun Liu and Dr. Sadegh Sabouri for discussing the idea, and to Diego Morra and Michele Tufano for searching in the Italian language.


\bibliographystyle{model2-names}
\bibliography{library}

\end{spacing}
\end{document}